\input{aipcheck}

\documentclass[
    ,final            
  ]
  {aipproc}

\layoutstyle{8x11single}

\begin{document}

\title{Many-Body Physics on the Border of Nuclear Stability}

\author{Alexander Volya}{
  address={Department of Physics, Florida
State University, Tallahassee, FL 32306-4350, USA}
}

\author{Vladimir Zelevinsky}{
  address={National Superconducting
Cyclotron Laboratory and Department of Physics and Astronomy,
Michigan State University, East Lansing, MI 48824-1321, USA}
}

\begin{abstract}
A brief overview is given of the Continuum Shell Model, a novel
approach that extends the traditional nuclear shell model into the
domain of unstable nuclei and nuclear reactions. While some of the
theoretical aspects, such as role and treatment of one- and
two-nucleon continuum states, are discussed more in detail, a
special emphasis is made on relation to observed nuclear
properties, including definitions of the decay widths and their
relation to the cross sections, especially in the cases of
non-exponential decay. For the chain of He isotopes we demonstrate
the agreement of theoretical results with recent experimental
data. We show how the interplay of internal collectivity and
coherent coupling to continuum gives rise to the universal
mechanism of creating pigmy giant resonances.
\end{abstract}

\maketitle


\section{Continuum Shell Model}
Recent progress in physics of weakly bound nuclei and other
marginally stable mesoscopic systems requires an adequate
many-body theory which would combine the bound states and
continuum in a consistent description. Below we show how it can be
done. Our formalism of continuum shell model (CSM) follows closely
the projection method by Feshbach \cite{feshbach58}. The many-body
states of the system are separated into subspaces ${P}$ and ${Q}$,
and the Hamiltonian consists of three parts,
$H=H_{QQ}+H_{PP}+H_{PQ}.$ The {\it internal} space ${P}$ contains
shell model (SM) states with all particles on bound
single-particle orbitals, for example $N$-nucleon $m$-scheme
Slater determinants, labelled as $|1\rangle $. The {\it external}
space ${Q}$ contains all states $|c;E\rangle$, labelled by energy
$E$ and channel $c$, with some particles in the continuum. These
wave functions asymptotically behave as outgoing spherical
(Coulomb) waves with respect to one or several single-particle
arguments. We solve the Schr\"odinger equation
$H|\alpha;E\rangle=E|\alpha;E\rangle$ within the total ${P}+{Q}$
space where an eigenstate $\alpha$ is a superposition
\begin{equation}
|\alpha;E \rangle = \sum_1 \alpha_1 (E)|1 \rangle + \sum_c \int
dE' \,\alpha_c(E';E)|c; E' \rangle,                   \label{2}
\end{equation}

Elimination of external states leads to the effective
energy-dependent Hamiltonian that acts within $P$ space,
\begin{equation}
{\cal H}(E)=H_{PP}+\Delta (E) -\frac{i}{2}\,W (E)\,. \label{heff}
\end{equation}
Here $\Delta(E)$ is a self-energy term coming from virtual
excitations into the continuum and $W(E)$ stands for the
explicitly non-Hermitian term due to the open decay channels,
\begin{equation}
\langle 1 |\Delta (E)|2\rangle = \sum_{c} \,P \int dE' \,
\frac{{A^c_1(E')}^*\, {A^c_2(E')}}{E-E'},  \quad
\langle 1 | W(E)|2\rangle = 2\pi  \sum_{c\,({\rm open})}
{A^c_1(E)}^*\,{A^c_2 (E)}  .                 \label{3}
\end{equation}
The amplitude of coupling to the continuum, $A^c_1(E)$, is
introduced as $\,A_1^c(E)=\langle c;E |H_{QP}| 1 \rangle\,$. For
Eqs. (\ref{3}) it is assumed for simplicity that the ${P}$ and
${Q}$ spaces are orthogonal, and that for the continuum
$H_{QQ}|c;E\rangle=E|c;E\rangle$. Thus, given the effective
Hamiltonian (\ref{heff}), the internal solution can be found as
\begin{equation}
\sum_2 {\cal H}_{12}(E) \alpha_2(E)={\cal E}\alpha_1(E),
\label{heff1}
\end{equation}
then, using the internal amplitudes $\alpha_1(E)$, the continuum
part can be restored as
\begin{equation}
\alpha_c(E';E)=\sum_1\,\alpha_1(E)\, \frac{A^c_1(E')}{E-E'+i0}\,.
\end{equation}
The energy dependence of the effective Hamiltonian (\ref{heff}) in
this formalism emphasizes the non-exponential nature of decay
processes. The experimental observables can be found through the
scattering matrix that is expressed via the same Hamiltonian:
\begin{equation}
S^{cc'}(E)=\exp(i\xi_c)\,\left\{\delta^{cc'}-\sum_{12}A^{c\ast}_{1}(E)\,
\left(\frac{1}{E-{\cal H}(E)}\right)_{12}A^{c'}_{2}(E)\right\}
\exp(i\xi_{c'}),                                       \label{scatt}
\end{equation}
where $\xi_c$ is a smooth scattering phase in channel $c$ coming
from features not accounted in the model Hamiltonian.

Equation (\ref{heff1}) has no solutions with real eigenvalues
${\cal E}$. This is not surprising since stationary intrinsic
states cannot be matched by outgoing continuum states in Eq.
(\ref{2}). However, solutions do exist in the complex energy plane
$E\rightarrow {\cal E}$ as Gamow quasistationary states
\cite{gamow28}, where resonance energies are taken as $E_{\rm
res}={\rm Re}({\cal E})$ and widths are identified as $\Gamma_{\rm
res}=-2 {\rm Im}({\cal E})$. The related non-Hermitian eigenvalue
problem is discussed in \cite{siegert39,berggren68,berggren96}.
This solution for the resonance lies in the foundation of the
alternative approach, so-called {\sl Gamow Shell Model}
\cite{michel03}, where the solution of the non-Hermitian
eigenvalue problem finds poles of the scattering matrix
(\ref{scatt}). The realization of this approach, although well
formulated mathematically, is complicated by the difficulty of
analytical continuation into the complex plane. In general, the
direct solution of Eq. (\ref{heff1}) leads to numerous unphysical
roots \cite{PRC67}, unless cuts in the complex momentum space are
made that separate physical and unphysical regions. The {\sl
Breit-Wigner formalism} gives an alternative definition of a
resonance \cite{breit36}, while avoiding transition into the
complex energy plain. Here both effective Hamiltonian and
scattering matrix are evaluated only on the real energy axis. The
resonance energy is determined from the condition that real part
in the denominator of scattering matrix vanishes which is
equivalent to finding $E_{\rm res}$ as $E_{\rm res}= {\rm
Re}({\cal E})$, where ${\cal E}=E_{{\rm res}}-(i/2)\Gamma_{{\rm
res}}$ is a complex eigenvalue of ${\cal H}(E_{\rm res})$. There
is yet another definition of a resonance that is closely related
to the experimental cross section when, from a phase shift
$\delta$ in a given channel, the resonance width is defined as the
phase shift derivative which at the resonance produces the maximum
time delay $\tau$,
\begin{equation}
\tau=\frac{2}{\Gamma}=\left . \frac {d\delta(E)} {d E} \right
|_{E=E_{{\rm res}}}, \quad \left . \frac { d^2\delta_l(E)} {d E^2}
\right |_{E=E_{\rm res}}=0  \,.             \label{resdef3_1}
\end{equation}
Having defined the {\it resonance region} as the energy region of
size $\Gamma$  around the resonance energy, all of the above
definitions agree if the two conditions are fulfilled: (1) the
resonance regions do not overlap; (2) the energy dependence of the
effective Hamiltonian in each of the resonance regions is weak and
can be ignored. In the realistic cases, when the density of states
is high and these conditions are violated, one has to be specific
and identify the definition of resonances and widths. In this work
we use a Breit-Wigner definition for the width. It is also
important that the CSM can be used for calculating the cross
sections which leads to direct comparison with experiment. In that
case, in contrast to a traditional SM, the large-scale matrix
diagonalization is substituted by the matrix inversion needed for
Eq. (\ref{scatt}).

\section{Application to weakly-bound nuclei}
In the following application of the CSM we assume that the
Hermitian part of the effective Hamiltonian $H_{PP}+\Delta$ is
given by the phenomenological SM Hamiltonian and limit $Q$-space
to one and two particles in the continuum. Phenomenologically
adjusted interactions include corrections from virtual excitations
to the continuum $\Delta$. Energy dependence of $\Delta(E)$ is
ignored since, unlike the non-Hermitian term $W$, $\Delta$ is not
sensitive to thresholds. This choice of the interaction assures
that, for the states below decay thresholds, coupling to continuum
disappears, $W=0$, and the results of the conventional SM are
exactly reproduced.

We assume that the mean field Hamiltonian selected here as a
Woods-Saxon potential is responsible for single-particle widths
while the additional schematic two-body interaction discussed
below is introduced in $H_{PQ}$ to mediate direct two-body decay
processes. With $b^\dagger(\epsilon)$ defined as a single-particle
creation operator in the continuum state with energy $\epsilon$, a
one-particle channel with quantum numbers $j$, energy of the
particle $\epsilon_{j}$, and the residual nucleus in the state
$|\alpha;N-1\rangle$, so that $E=E_{\alpha}+\epsilon_{j}$, is
asymptotically labelled as
$|c\rangle=b^{\dagger}_{j}(\epsilon_{j})|\alpha;N-1\rangle$. The
two-body channel can be expressed similarly, $|c\rangle =
b_j^\dagger(\epsilon) b^\dagger_{j'}(\epsilon') |\alpha;N-2
\rangle.$ The first contribution from the continuum comes through
a single-particle decay amplitude
\begin{equation} A^c_1(E_\alpha+\epsilon_j)=
\sum_\nu a^j_\nu(\epsilon_j)\,\langle
\alpha;N-1|b_\nu|1;N\rangle,                       \label{7p}
\end{equation}
where the reduced single-particle amplitudes are
\begin{equation}
a^j_{\nu}(\epsilon)=\langle 0|b_{\nu}(\epsilon)\, H_{PQ}
\,b^\dagger_j|0 \rangle,                          \label{8}
\end{equation}
and the operators $b^{\dagger}_\nu$ create particles on the
discrete SM orbitals. If the valence space is small, each
single-particle state $\nu$ is uniquely identified by spin-isospin
quantum numbers $j$. As a result, the contribution to the
effective Hamiltonian from one-body decay is
\begin{equation}
\langle 1 | W(E)|2\rangle = 2\pi \delta_{1 2} \sum_{c\,({\rm
open})} |a^{j}(\epsilon_j)|^2 |\langle
\alpha;N-1|b_j|1;N\rangle|^2,                   \label{9}
\end{equation}
where $E=\epsilon_j+E_\alpha$.

Being in general a many-body operator, $W$ is effectively reduced
to a single-particle form in some important cases. For example,
far from thresholds one can ignore the energy dependence and use
the closure to simplify the summation in (\ref{9}),
\begin{equation} \langle 1 | W(E)|2\rangle = 2\pi \delta_{1 2}
\sum_{j} |a^{j}|^2\, |\langle 1;N|b^\dagger_j b_j|1;N\rangle|^2.
                                              \label{10}
\end{equation}
$W$ than becomes a one-body operator that assigns a width $
\gamma_j=2 \pi \left |a^{j}\right |^2   $ to each unstable
single-particle state $j=\nu$ and can be combined with the shell
model Hamiltonian via complex single-particle energies
$e_\nu=\epsilon_\nu-i \gamma_\nu/2.$ The single-particle
interpretation of the width is possible when the decay process is
not associated with a significant change in the structure from
parent to daughter nucleus. The coupling amplitude
$a^j(\epsilon)$, Eq. (\ref{8}), is calculated numerically through
a one-nucleon scattering problem. Within few MeV above decay
thresholds, the energy-dependence of $a^j(\epsilon)$ to a good
accuracy can be parameterized by the power law consistent with the
energy scaling of the width as $\Gamma(\epsilon)\sim
\epsilon^{l+1/2}$ where $\epsilon$ is the energy above the
threshold and $l$ is the partial wave of the decay channel.

With the one-body interaction used for single-particle decays, the
admixture from the two-particle continuum can only appear as a
second order contribution,
\begin{equation}
A^c_1(E)= \sum_\beta a^j(\epsilon) a^{j'}(\epsilon') \left
(\frac{\langle \alpha|b_{j}|\beta \rangle\,\langle \beta
|b_{j'}|1\rangle}{E-E_\beta-\epsilon}+ \frac{\langle
\alpha|b_{j'}|\beta\rangle\,\langle \beta|b_{j}|1\rangle}
{E-E_\beta-\epsilon'}\right )\,,               \label{sequential}
\end{equation}
that proceeds through an intermediate state $\beta$ of the
nucleus with $N-1$ particles ({\it sequential} two-body decay).
The contribution from this decay to matrix $W$ is
\begin{equation}
\langle 1|W(E)|2\rangle =
2\pi \sum_{\alpha,j,j'}\int\, d\epsilon\, d\epsilon'\,
\delta(E-E_\alpha-\epsilon-\epsilon')\, {A^c_1(E)}^* A^c_2(E) \,.
                                         \label{sequential2}
\end{equation}
The integration over single-particle energies may contain poles
corresponding to open one-body decays. We concentrate here on the
principal value part that represents off-shell processes. Eqs.
(\ref{sequential}) and (\ref{sequential2}) allow for additional
simplifications in the near-threshold region. The overall width
behaves as $\Gamma\sim E_k^{2+l+l'}$ where $E_k$ is the total
available kinetic energy.

In contrast to sequential decay, the amplitude for the {\it
direct} two-body transition is mediated by the two-body
interaction in the $H_{PQ}$ part of the Hamiltonian. To describe
this process, a pair amplitude is introduced,
\begin{equation}
A^c_1(E)=a^{(L)}(\epsilon_1, \epsilon_2)\,\, \langle
\alpha;N-2|p_L|1;N \rangle\,,              \label{14}
\end{equation}
where the operator $p_L=\{b_\nu \otimes b_{\nu'}\}_L$ removes a
pair coupled to quantum numbers $L$. Unlike in the sequential
case, direct pair emission conserves only the quantum numbers of
the pair. Near thresholds we parameterize direct decay amplitude
$a^{(L)}(\epsilon_1, \epsilon_2)$ with the appropriate energy
dependence that reflects the phase space kinematics of the
free-body final state.

\begin{figure}[t]
  \includegraphics[width=.8\textwidth]{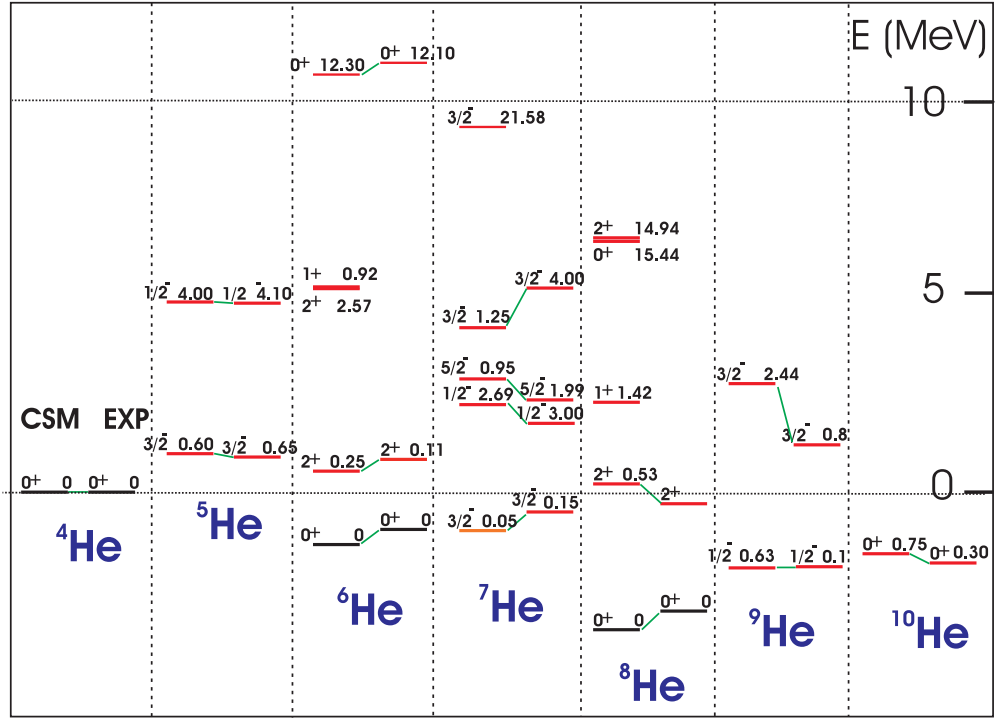}
  \caption{\label{hes}CSM calculation for He isotopes. For each
  isotope, the CSM result is on the left and experimental value
  is on the right, energy of each state or resonance is not shown
  but the energy scale is given on the vertical axis. For each
  state spin, parity and the decay width in MeV are displayed. }
\end{figure}
As an example, we consider the $p$-shell chain of helium isotopes
from $^4$He to $^{10}$He. The internal space consists of two
single-particle levels, $p_{3/2}$ and $p_{1/2}$. The SM
interaction and single-particle energies are defined in
\cite{cohen65,stevenson88}. For the one-body part of $H_{QQ}$ and
$H_{PQ}$ we use the Woods-Saxon potential adjusted for
single-particle states in $^5$He. Even at several MeV above
threshold, the single-particle widths for the $p$-wave decay from
$p_{3/2}$ and $p_{1/2}$ levels to a good accuracy can be described
by the parameterizations $\gamma_{3/2}(\epsilon)=0.608
\epsilon^{3/2}$ and $\gamma_{1/2}(\epsilon)=0.3652 \epsilon^{3/2}$
(all energies in MeV). The sequential two-body decay is computed
using Eq. (\ref{sequential}) with the near-threshold approximation
and the above assumption for the energy dependence of
single-particle amplitudes. The calculation with just one-body
part in $H_{PQ}$ is inadequate emphasizing the need for the direct
two-body decay. This decay is introduced only for pair emission
with total angular momentum $L=0$. It is assumed that all $L=0$
neutron pairs couple to the continuum with the same decay
amplitude $a^{(L=0)}(\epsilon_1, \epsilon_2)$ parameterized as $
a^{(L=0)}(\epsilon_1, \epsilon_2)= {|\epsilon_1+\epsilon_2|}/
{(3.0 \sqrt{2\pi})} . $ The numerical constant 3.0, the only
parameter of the model, is introduced to describe the unknown
strength of the residual two-body interaction needed for two-body
decay of $^{6}$He. With parameters of the model being identified,
the full solution proceeds along the chain of isotopes starting
from $^4$He, so that at each step the relevant states in the
daughter systems are known. For each state in an $N$-particle
system, the solution is iterative, energy $E$ is used to determine
possible open decay modes into $N-1$ and $N-2$ daughters, then the
non-Hermitian part $W(E)$ is constructed, and the full effective
Hamiltonian is diagonalized leading to a new complex energy
eigenvalue. For each state in the system this procedure is
repeated until convergence.

Figure \ref{hes} shows the level scheme obtained with the CSM
calculation for He isotopes along with experimental data. Beyond
good overall agreement, additional fine details can be noted that
are in close relation to recent experimental data
\cite{korsheninnikov99,rogachev03,rogachev04}. For $^6$He the
$2^+_{1}$ state is well described by the sequential two-body decay
into $^4$He. However, the description of the broad first excited
$0^+$ state requires direct two-body decay, in agreement with the
coherent $L=0$ pair-excitation structure of that state
\cite{PRC65}. For $^7$He, the CSM confirms the peculiar nature of
the 5/2$^-$ state noticed in the experiment
\cite{korsheninnikov99}: this state decays to the excited $2^+$
state rather than to the ground state of $^6$He. This is not the
case for the neighboring state $1/2^-$ that mainly decays to the
ground state. Theory and very recent experimental results
\cite{rogachev04} are in good agreement regarding spectroscopic
factors and branching ratios. Additional CSM studies of realistic
nuclear cases (the chain of oxygen isotopes) and related
discussions can be found in Ref. \cite{volya_prl}.

\section{Low energy branch of giant resonance}
The CSM presented here combines structure and reaction physics on
the level that goes beyond any perturbative treatment; the
unitarity of the scattering matrix is fulfilled automatically. The
interesting effects related to the nature of many-body states
embedded in the continuum can be noted in the above example of He
isotopes. Compared to the conventional SM, the centroids of
low-lying decaying states are shifted into continuum, see Ref.
\cite{volya_prl}; in addition, the coupling to the continuum
affects mixing of states and their spectroscopic factors.
\begin{figure}
\begin{minipage}[b]{0.40\linewidth}
The drawing schematically shows the spectrum of particle-hole
excitations on the right. Due to multipole interaction, the giant
collective state (GR) is repelled up in energy. At the same time,
coupling to the common particle continuum creates super-radiant
(SR) collectivity visible through the particle scattering
reaction. The cross section of such a reaction is sketched on the
left. Factorized forms of the real interaction and of the coupling
to continuum assure collectivity in both GR and decay modes. The
dynamics are determined by two multidimensional vectors ${\bf
d}=\{d_{n}\}$ and ${\bf A}=\{A_{n}\}$, see text. The multipole
interaction shifts the GR accumulating the multipole strength by
$\Omega=\lambda {\bf d}^{2}$ {\sl along the real energy axis} from
the unperturbed centroid $\bar{\epsilon}$. The continuum
interaction shifts the SR state accumulating the decay width
{\sl along the imaginary energy axis}.
\end{minipage}\hfill
\hskip 0.8 cm
\begin{minipage}[b]{0.38\linewidth}
\includegraphics[height=.3\textheight]{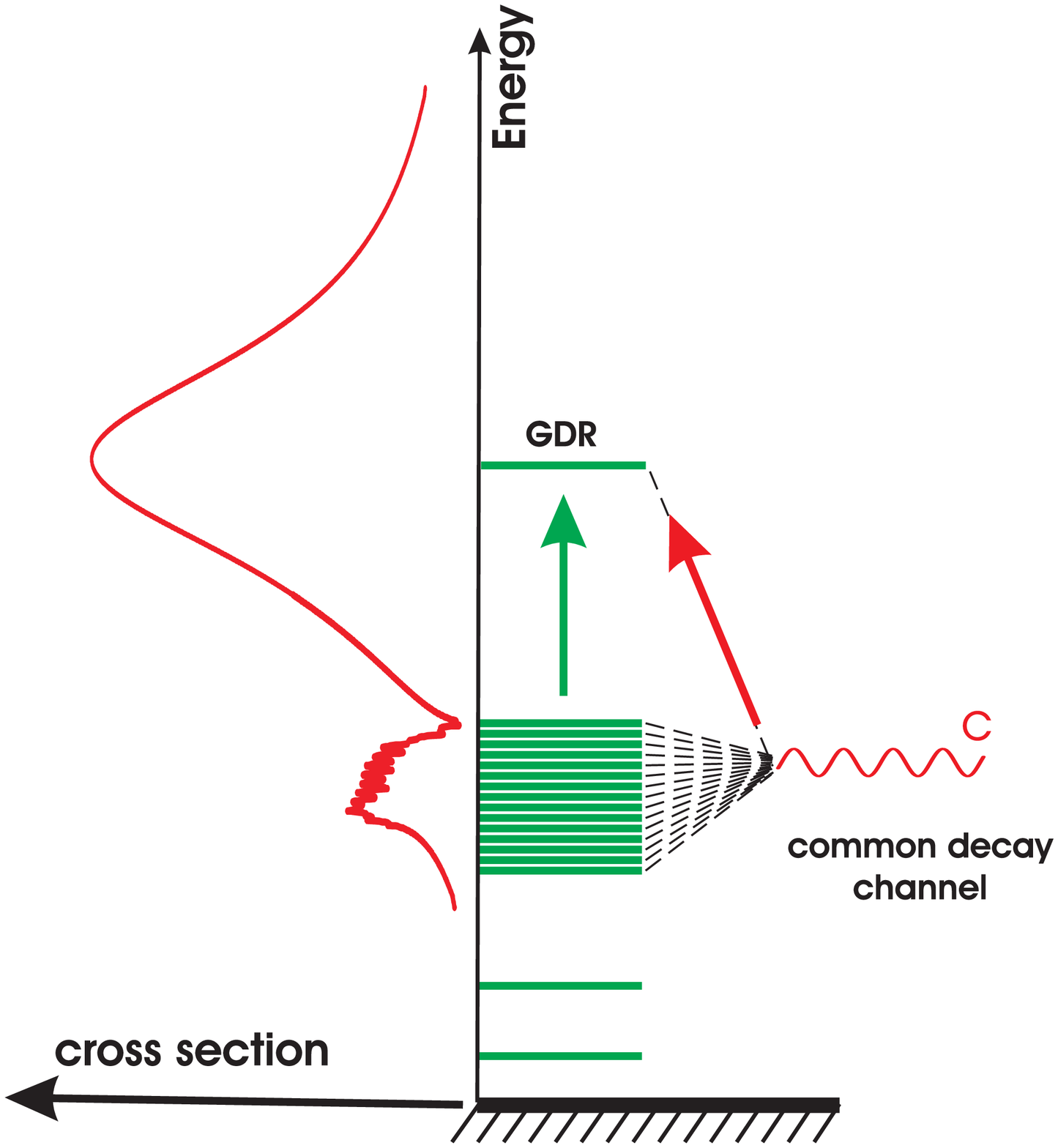}
\vspace{1 cm}
\end{minipage}
\caption{\label{pigmy}}
\end{figure}

It was first noted by Dicke \cite{dicke54} that strong coupling of
two-level atoms to a common radiation field creates {\sl
superradiant} (SR) coherence. The phase transition to the
superradiant regime was found to be a universal phenomenon in
systems with overlapping resonances ranging in nature from
molecular \cite{JOB5,pavlov88,flambaum96,persson00} to hadronic
(pentaquark \cite{pentaquark}). The universality of the SR
mechanism \cite{sokolov89} comes from the factorizable nature of
continuum coupling in Eq. (\ref{3}). With a few continuum
channels, the rank of the matrix $W$ is much smaller than the
dimension of the intrinsic space. Strong coupling to continuum
reorganizes the system in favor of creation of few SR states while
the remaining states become long-lived being nearly orthogonal to
decay.

Giant resonance (GR) in the response function of the nucleus to a
multipole excitation appears as a result of the coherent coupling
of particle-hole excitations. In many cases excitation energies of
particle-hole configurations lie above the particle separation
energy which may lead to another collectivization towards particle
emission. We address this interplay in the simplified example
\cite{sokolov90,sokolov_rotter97} assuming the effective
Hamiltonian
\begin{equation}
{\cal H}_{12}=\epsilon_{1}\delta_{12} +\lambda d_{1}d_{2}
-\frac{i}{2}A_{1}A_{2}                      \label{6}
\end{equation}
that contains unperturbed energies, real multipole interaction and
interaction through the continuum, see Fig. \ref{pigmy} and the
caption for additional discussion.
\begin{figure}
  \includegraphics[width=.75\textwidth]{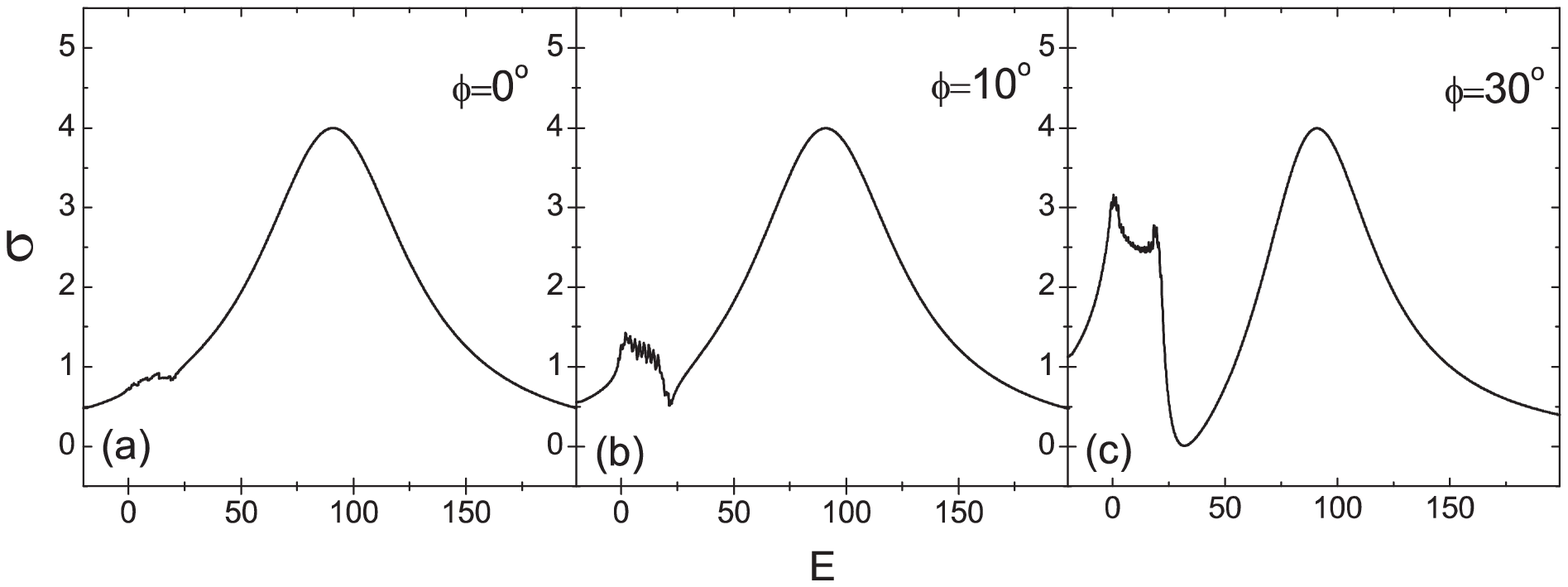}
  \caption{\label{pigmy2}Particle scattering cross section showing the
excitation of GR for different angles between GR and SR collectivities.}
\end{figure}
The interplay of the collective effects depends on the ``angle"
$\phi$ between the vectors ${\bf d}$ and ${\bf A}$. In the
degenerate case, $\epsilon_{n}=\epsilon$, the reaction amplitude
is
\begin{equation}
T(E)=\frac{(E-\epsilon-\Omega)\Gamma+\lambda ({\bf A}\cdot{\bf
d})^{2}}{(E-\epsilon-\Omega)[E-\epsilon+(i/2)\Gamma]+(i/2)\lambda
({\bf A}\cdot{\bf d})^{2}}.                    \label{7}
\end{equation}

In the limits of {\sl parallel} internal and external couplings,
$\phi=0^\circ$, the SR and GR collectivizations are coherent, and
the experiment will reveal the shifted ``{\sl Giant Dicke
resonance}" with full strength and full width $\Gamma={\bf
A}^{2}$. This is the case when particle scattering would be most
efficient in excitation of the GR. The opposite case of orthogonal
couplings, $\phi=90^\circ$, will result in a {\sl dark} collective
state with no access to the continuum. Fig. \ref{pigmy2} shows a
numerical calculation for non-degenerate intrinsic states when the
strength and the width are both shared between the displaced GR
and the SR in the region of unperturbed excitation energies. With
the increasing angle $\phi$ the coupling with continuum becomes
more fragmented and less effective in driving the GR; the missing
strength develops a low-energy branch revealing the universal
mechanism for creating the so-called {\sl pigmy giant resonance}.

\section{Final notes and acknowledgments}
We emphasized some of the very recent developments along the road
from the nuclear structure and stable nuclei to nuclear reactions
and continuum. We were able to highlight only a handful of
examples and techniques, a tiny part of a huge modern-day effort.
The main message is that the CSM today is no longer a theoretical
concept, it is a powerful practical tool capable of describing
many-body systems on the border of stability with an impressive
quality.\\
\\
The authors acknowledge support from the U. S. Department of
Energy, grant DE-FG02-92ER40750; Florida State University FYAP
award for 2004, and National Science Foundation, grant
PHY-0244453. Useful discussions with B.A. Brown and G.V. Rogachev
are highly appreciated.
\bibliographystyle{aipproc}   


\end{document}